\documentclass[journal]{IEEEtran}

\usepackage{amsmath}
\usepackage{graphicx}
\usepackage{subfigure}
\usepackage{multicol} 

\usepackage{booktabs}  
\usepackage{threeparttable} 
\usepackage{multirow}
\usepackage{amssymb}

\usepackage{color}

\begin{document}
%
\title{ Neural System Identification with Spike-triggered Non-negative Matrix Factorization}

%
%
%

\author{
Shanshan Jia, Zhaofei Yu, Arno Onken, Yonghong~Tian, 
Tiejun~Huang,  
Jian~K.~Liu
	
\thanks{S. Jia, Z. Yu, Y. Tian, T. Huang are with the National Engineering Laboratory for Video Technology, School of Computer Science and Technology, Peking University, Beijing 100871, China, and also with Peng Cheng Laboratory, Shenzhen 518055, China. e-mail: (jssll00@126.com, yuzf12@pku.edu.cn, yhtian@pku.edu.cn, tjhuang@pku.edu.cn).} 
\thanks{A. Onken is with the Institute for Adaptive \& Neural Computation, School of Informatics, University of Edinburgh, Edinburgh, UK. e-mail: (aonken@inf.ed.ac.uk).}
\thanks{J. K. Liu is with the Centre for Systems Neuroscience, Department of Neuroscience, Psychology and Behaviour, University of Leicester, Leicester, UK. e-mail: (jian.liu@leicester.ac.uk).}

}


%

\maketitle

\begin{abstract}
Neuronal circuits formed in the brain are complex with intricate connection patterns. Such a complexity is also observed in the retina as a relatively simple neuronal circuit. A retinal ganglion cell receives excitatory inputs from neurons in previous layers as driving force to fire spikes. Analytical methods are required that can decipher these components in a systematic manner. Recently a method termed spike-triggered non-negative matrix factorization (STNMF) has been proposed for this purpose. In this study, we extend the scope of the STNMF method. By using the retinal ganglion cell as a model system, we show that STNMF can detect various computational properties of upstream bipolar cells, including spatial receptive field, temporal filter, and transfer nonlinearity. In addition, we recover synaptic connection strengths from the weight matrix of STNMF. Furthermore, we show that STNMF can separate spikes of a ganglion cell into a few subsets of spikes where each subset is contributed by one presynaptic bipolar cell. Taken together, these results corroborate that STNMF is a useful method for deciphering the structure of neuronal circuits.
\end{abstract}


%
\IEEEpeerreviewmaketitle

\section{Introduction}
\begin{figure}[t!h]
	\begin{center}
		\includegraphics[width=\columnwidth]{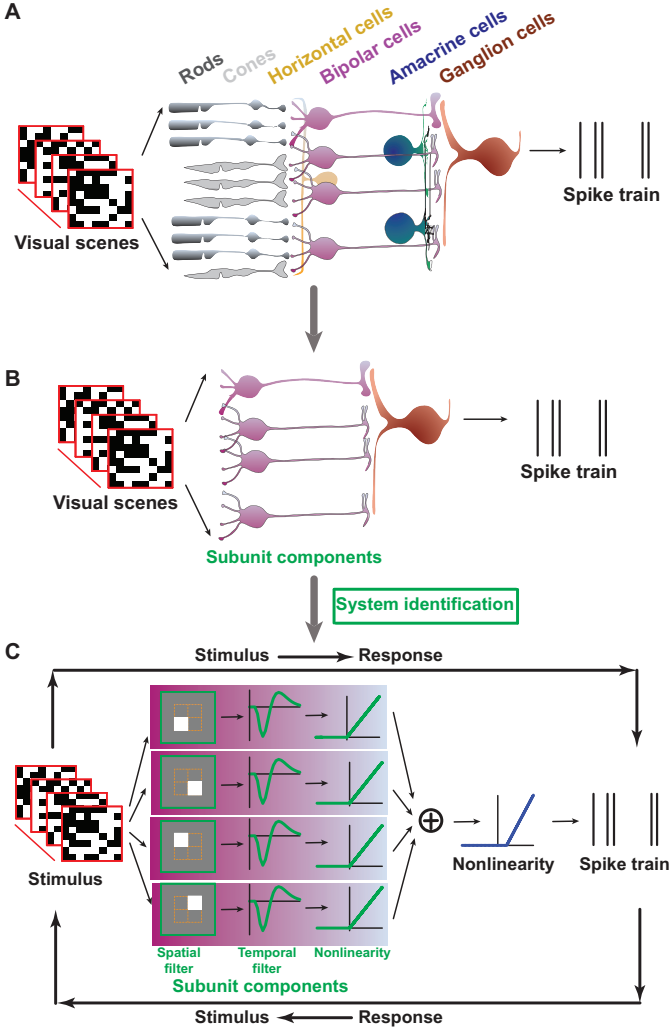}
	\end{center}
	\caption{ Illustration of retinal neuronal network and ganglion cell model. (A)~Illustration of retinal network. Light signals come into the retina from photoreceptors (rods and cones), transfer to bipolar cells, and then send the output as spikes of the ganglion cell. In between, there are horizontal cells and amacrine cells as inhibitory modulations. (B) A minimal neural network of one ganglion cell consisted of a few bipolar cells as subunit components. System identification is to uncover properties of these subunit components. (C)~Model illustration. Stimuli are a sequence of random black-white images in 8x8 pixels. The network has four subunits, where each one has a spatial filter covering a 2x2 region indicated by the white color (1 in white, 0 everywhere else), temporal OFF filter and threshold-linear nonlinearity. Summation of the output of each subunit is pooling to a readout unit, passes the final nonlinearity, and generates the activity as a spiking probability to determine actual spikes according to Poisson process.  
    } \label{fig:retina}
\end{figure}

%
%
%
%
\IEEEPARstart{N}{euronal} circuits in the brain are highly complex. Even for the retina, a relatively simple neuronal circuit, the underlying structure and, in particular, its functional characteristics are still not completely understood. However, the retina serves as a typical model for both deciphering the structure of neuronal circuits~\cite{helmstaedter2013connectomic, zeng2017neuronal, marc2013retinal, seung2014neuronal, sanes2015types, demb2015functional} and testing novel methods for neuronal coding~\cite{Chichilnisky2001a, Pillow2008Spatio,McFarland2013, Yu2020, Zhang2020}. The retina consists of three layers with photoreceptors, bipolar cells, and ganglion cells together with inhibitory horizontal and amacrine cells in between as illustrated in Fig.~\ref{fig:retina}(A). The ganglion cells (GCs), as the only output neurons of the retina, send visual information via the optic tracts and thalamus to cortical areas for higher cognition. Each ganglion cell receives input from a number of excitatory bipolar cells (BCs) as driving force to generate spikes (Fig.~\ref{fig:retina}(B)). 

Due to clear input-output relation, the retina is well suited for studying encoding/decoding of stimulus (visual optical image) with neuronal responses (spikes in retinal GCs). For the purpose of system identification, characterizing its neuronal circuit is not trivial. However, most methods for deciphering neuronal circuits rely on experimental techniques. Traditionally, one can detect the connection between neurons with single or multiple electrodes~\cite{Song2005, Jiang_2013}. With the advancement of experimental techniques, large-scale multielectrode array and calcium imaging can simultaneously record hundreds or thousands of cells. Therefore, a systematic method for analyzing these cells is highly desirable.

A recent work proposed such a method, termed spike-triggered non-negative matrix factorization (STNMF), to analyze the underlying structural components of the retina~\cite{Liu2017}. Non-negative matrix factorization (NMF) has been proposed to capture the local structure of a given dataset~\cite{Lee_1999}. It is widely used in computer vision~\cite{Zhao_2016,ye2015multitask}, signal processing~\cite{Kwon_2015,guan2012nenmf,gao2014machine}, machine learning~\cite{pei2014automated,blumensath2016directional,wang2011fast}, gene expression~\cite{Devarajan_2008}, and neuroscience~\cite{gold2011comparing, maruyama2014detecting, Beyeler_2016, pnevmatikakis2016simultaneous, zhou2018efficient}. The ability to learn local parts from the whole dataset is further improved by sparseness constraints~\cite{Hoyer_2004, Eggert_2004}. Such a sparse coding is naturally related to the receptive field structure of sensory neurons which is typically found in visual system~\cite{Olshausen1996, Hoyer_2003}. 

In the recent study~\cite{Liu2017}, by analyzing the spikes recorded from the retinal GCs, STNMF was shown to identify physical locations of subunit bipolar neurons of the previous layer that are pooling to a target retinal GC. Here we significantly extend this approach to demonstrate how STNMF can be used for characterizing various functional properties of the retinal circuit. It is difficult to demonstrate the power of STNMF for a biological neuronal circuit, even the retina, as there are many unknowns due to the limitations of current experimental techniques for measuring a complete map of the retina. Therefore, in this study, we first use a clearly defined minimal network model as proof of principle to explain STNMF, and then demonstrate it with the retinal GC data. 

The rest of this paper is structured as follows. First, we explain the workflow of STNMF as a general framework for system identification of neural network for a modeled retinal ganglion cell in Sec.~\ref{method}. Sec.~\ref{result} shows a complete picture of neural network components, including synaptic subunit structures, synaptic connections, and their weights between presynaptic BCs and postsynaptic GC. Then we demonstrate a novel feature of STNMF for classification of all the spikes of a GC into a few subsets of spikes, where each subset of spikes is mainly contributed by one presynaptic BC. 
When applying STNMF to biological data of retinal ganglion cells, similar results are found as for the artificial data with known ground-truth. For each retinal ganglion cell, STNMF finds a set of presynaptic BCs together with their contributed spikes. The paper concludes with a summary and discussion in Sec.\ref{discussion}.


\section{ Methods }\label{method}

For a biological neuronal circuit, even the retina as illustrated in Fig.~\ref{fig:retina}(A), as there are many unknowns. Thus, we use a clearly defined minimal network model of the retinal ganglion cell as proof of principle to explain the framework STNMF as a method of system identification for neural network.
\subsection{Ganglion cell model}
A simulated ganglion cell illustrated in Fig.~\ref{fig:retina} is modeled by a typical linear-nonlinear model that has been shown to capture biological retinal neuronal responses~\cite{ McFarland2013,Liu2017,gollisch2008modeling}. The model cell has four excitatory subunits with a size of 2 x 2 pixels each. The setup of the model is equivalent to a neural network with two layers: an input layer with four subunits and an output layer with one single readout unit. Inhibitory neurons are not modeled here since they are barely triggered to see the effect on the receptive field of the ganglion cell under the stimulation condition of white noise.

Input stimuli are given by a sequence of random binary black or white checkers. Similar to those filters in the visual cell, each subunit has a static spatial filter and a temporal filter. The different subunits have different spatial locations such that each subunit can only ``read" input stimuli at one specific location, but ignores other parts outside this location. After input stimuli are convolved by these subunits with spatial and temporal filters, the filter outputs then pass a rectification stage in the form of a threshold-linear nonlinearity. The outcomes of all subunits are summed up with a weight for each subunit and pooled to the output unit. Finally, the summation is rectified by another threshold-linear nonlinearity with a higher positive threshold to get the final output. Note that since the summation is already positive, a higher threshold is needed to reduce baseline activity and generate spare spiking activity. In the end, a spike train is sampled from a Poisson process. 

Note that the current model is implemented for OFF-type retinal ganglion cells with subunits having OFF polarity, i.e., the linear filter (as a multiplication of spatial and temporal filter) prefers the negative part of the stimulus image. One can simply tell the polarity by fixing the spatial filter to be positive (indicated in white, comparing to black-white stimulus in Fig.~\ref{fig:retina}(C)), and checking the dominant part (the first peak close to spiking time) of the temporal filter to be positive or negative. 

Such a model can be considered as a minimal network of ganglion cell consisted of four bipolar cells as subunit components. The recent study~\cite{Liu2017} used STNMF to identify physical locations of subunit bipolar cells that are pooling to a target retinal GC. However, no functional properties of the bipolar subunits were uncovered there. Here we use this model to demonstrate how STNMF can be used for characterizing functional properties of bipolar subunits, which includes spatial and temporal filters, nonlinearities, synaptic connections and strengths, and more importantly, subset of ganglion cell spikes contributed by each bipolar cell.

\subsection{Spike-triggered analysis}
The STNMF method is based on a simple and useful method for system identification in visual neuroscience, so-called ``spike-triggered average (STA)"~\cite{Chichilnisky2001a, vance2018bioinspired}, which is similar to the first order kernel in the Volterra/Wiener kernel series expansion~\cite{Sandler_2015}: 
 \begin{align} \label{eq:sta}
  r(t)  & =  \int_{\mathcal{R}} k(\tau) s(t-\tau) d \tau \nonumber \\
        & + \int_{\mathcal{R}^2} h(\tau_1, \tau_2) s(t-\tau_1) s(t-\tau_2)  d \tau_1 \tau_2 + \cdots .  \nonumber \\
 \end{align}
When stimuli are Gaussian, both kernels $k(\tau)$ and $h(\tau_1,\tau_2)$ can be estimated by reverse correlation. Specifically, for the $i$-th spike $r^i$ occurring at time $t_i$, one collects a segment of stimuli $s(\tau)^i = s(t_i - \tau)$ that preceded that spike, where the lag $\tau$ denotes the timescale of history, into an ensemble of spike-triggered stimuli $\{  s(\tau)^i \}$, then averages it over all spikes to get the STA filter $k(\tau) = \langle s(\tau)^i \rangle_i$. When the stimuli are spatialtemporal white noise, the 3D STA filter can be decomposed by singular value decomposition to get the principle temporal filter and spatial receptive field~\cite{Gauthier_2009}. 
An illustration of the spatial receptive field of the GC model obtained by STA is seen in Fig.~\ref{fig:model}. 

\subsection{Spike-triggered non-negative matrix factorization analysis}

\begin{figure}[t!h]
	\begin{center}
		\includegraphics[width=\columnwidth]{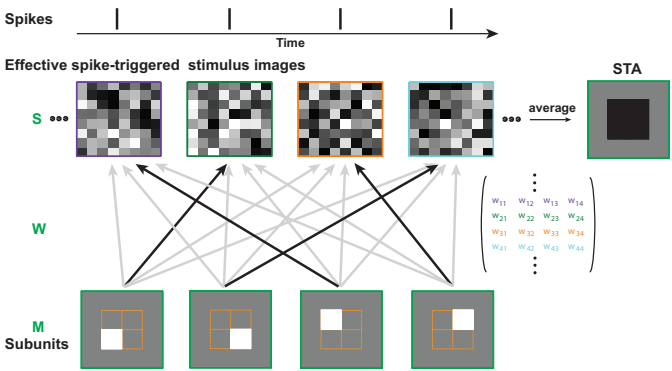}
	\end{center}
	\caption{ Illustration of STNMF analysis. For each spike, there is an effective spike-triggered stimulus image. Averaging of this ensemble yields a single STA filter. STNMF reconstructs this image ensemble by approximating with a set of modules and a matrix of weights such that one of the modules is strongly correlated to one of spikes/images indicated by stronger (black lines) or weaker (gray lines) weights. 
    } \label{fig:model}
\end{figure}

The procedure of spike-triggered non-negative matrix factorization analysis is similar to the one described in~\cite{Liu2017}. Briefly, to reduce computation costs for STNMF analysis, we first apply a pre-processing for the spike-triggered stimulus ensemble: for the $i$-th spike, the corresponding stimulus segment $s(\tau)^i$ is weighted averaged by the temporal STA filter $k_t$: $\bar{s}^i = s(\tau)^i \cdot k_t(\tau)$ such that time dimension $\tau$ is collapsed. This results in a single frame of stimulus image for the $i$-th spike, termed effective stimulus image $\bar{s}^i$. With the ensemble of effective stimulus images $S = \{  \bar{s}^i \}_i$ for all spikes as illustrated in Fig.~\ref{fig:model}, one can apply NMF directly, similarly as one would analyze a set of face images~\cite{Lee_1999}. 
Specifically, $S = (s_{ij})$ is a $N \times P$ matrix with indices $i=1,\cdots, N$ for all $N$ spikes, and $j=1,\cdots, P$ for all $P$ image pixels.  We used a semi-NMF algorithm~\cite{Ding_2010} such that 
 \begin{align} 
  S \approx W M 
 \end{align}
where weight matrix $W$ is $N \times K$, module matrix $M$ is $K \times P$, and $K$ is the number of modules. Both stimuli $S$ and weights $W$ can be negative, but modules $M$ are still non-negative.

The idea of semi-NMF can be understood from the perspective of clustering. One could consider the data matrix $S=(s_1,\cdots,s_P)$ as a collection of $P$ vectors as columns. Each vector $s_j$ is a sequence of effective stimulus images at a specific spatial location since the number of pixels is corresponding to the total space of an image. Suppose we have a $K$-means clustering on $S$ with cluster centroids $W = (w_1,\cdots, w_K)$. Thus each $w_k$ is a sequence of weights, in which each individual weight $w_{ik}$ is the strength between the $i$-th spike-triggered (effective) stimulus image and the module $k$. Larger $w_{ik}$ means stronger correlation between the $i$-th spike and the module $k$. Therefore, the matrix $W$ reflects connection weights between spikes and modules/subunits. Biologically, this is equivalent to the synaptic weight from a presynaptic neuron to a postsynaptic neuron. In this way, STNMF essentially becomes a clustering method by connecting those spikes generated by subunits with a set of modules such that each module/subunit contributes its corresponding spikes locally at a specific space as illustrated in Fig.~\ref{fig:model}.

If we let $M = m_{kj}$ denote the cluster indicators, i.e., $m_{kj}=1$, if $s_j$ belongs to cluster $k$, $m_{kj}=0$, otherwise. We can write the $K$-means clustering objective function as 
\begin{align} 
  F_{K-\mathrm{means}}  & =  \sum_{j=1}^P \sum_{k=1}^K m_{kj} \parallel s_j - w_{k}  \parallel^{2}_{2} = \parallel S - WM \parallel^{2}_{F},  
 \end{align}
where $\parallel v \parallel $ denotes the $L_2$ norm of a vector $v$ and $ \parallel A \parallel  $ denotes the Frobenius norm of a matrix $A$. The above objective can be alternatively considered as an objective function for matrix approximation. The difference is that $M$ is not binary but non-negative $M \in R^+ $. In addition, a sparseness constraint is added on the columns of $M$~\cite{Kim_2007} such that
 \begin{align} 
  F = \parallel S - W  M \parallel^{2}_{F} + \lambda \sum_{j=1}^{P}  \parallel  M_j  \parallel_1^2,
 \end{align}
where the sparsity parameter $\lambda=0.1$ throughout the current study, and $\parallel v \parallel_1  $ is the $L_1$ norm of a vector $v$. The minimization of $F$ was implemented as an alternating optimization of $W$ and $M$ based on the NMF Matlab toolbox~\cite{Li_2013}.

\begin{figure}[th]
	\begin{center}
		\includegraphics[width=\columnwidth]{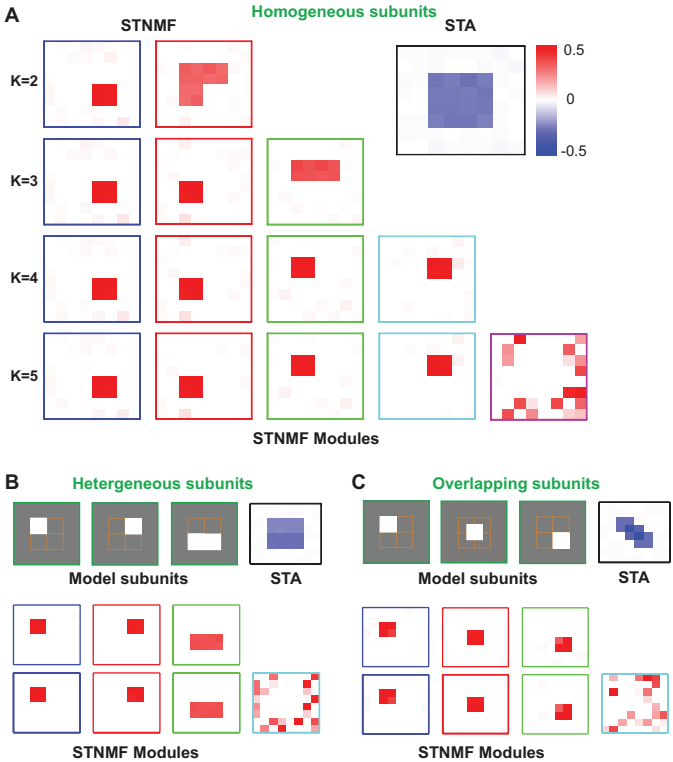}
	\end{center}
	\caption{ Exact subunit spatial filter revealed by STNMF. (A)~STNMF faithfully recovers the original subunits when $K=4$. (Left) STNMF modules with the parameter $K=2, 3, 4, 5$ when the subunits are homogeneous. (Inset) Receptive field as spatial STA. Color bar indicates positive (red) and negative (blue). For better visualization, STA is flipped as negative. (B)~Heterogeneous subunits by STNMF. (Top) Model subunits with small and big regions, but STA is similar. (Bottom) STNMF modules with $K=3,4$. (C)~Similar to (B) but with model subunits overlapped in space. } \label{fig:nmf}
\end{figure}

\section{Results}\label{result}

\subsection{Subunit filters revealed by STNMF} We set up a minimal model of the retinal GC as in Fig.~\ref{fig:retina}(C) in order to investigate how upstream BCs affect spiking activity of the target GC (see Methods for details). The GC model has four subunits as excitatory BCs that have spatial and temporal filters to compute the stimulus. The final spikes of GC are the only output of this model. With the input of stimulus images and the output of GC spikes, the question is how to do system identification to find these computational components used by the model.

Similar to typical experimental protocols~\cite{Chichilnisky2001a,Liu2017}, we used visual stimuli consisting of a sequence of white noise as black and white checkers randomly distributed in the space and time domains. Under this stimulation, the receptive field of GC can be computed from spiking response by a method named spike-triggered average (STA)~\cite{Schwartz2006} (see Methods).
This STA is equivalent to an average characteristic of GC, so the shape of STA is a combination of all subunits in space as shown in Fig.~\ref{fig:model}. Note that inhibitory neurons are not included in the model as the surround of receptive field is barely triggered under the white noise stimulation~\cite{Liu2017}.

However, computations are done by the subunits of the model in the first case. Extracting the detailed information of these subunits can be achieved by another recently proposed method, named spike-triggered non-negative matrix factorization (STNMF)~\cite{Liu2017}.
The framework of STNMF (see Methods) is illustrated in Fig.~\ref{fig:model}. In the previous study~\cite{Liu2017}, STNMF was shown to identify the spatial receptive field of subunits of modeled ganglion cells. Here we recap this finding and additionally show that the working principle of STNMF is to reveal the underlying nonlinear computation of the network.

\begin{figure}[th!]
	\begin{center}
		\includegraphics[width=\columnwidth]{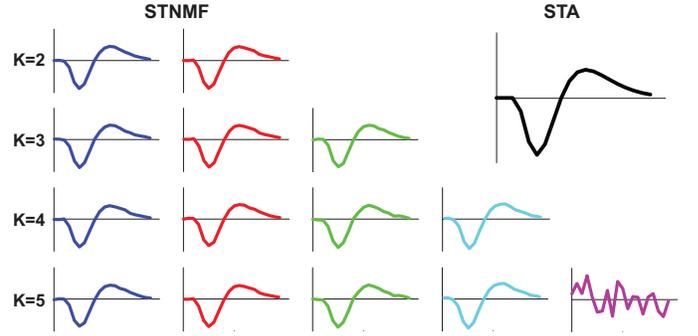}
	\end{center}
	\caption{ Recovered subunit temporal filter for different subunits with $K=2, 3, 4,5 $. (Inset) STA temporal filter. Same color as in Fig.~\ref{fig:nmf}(A). } \label{fig:tm1}
\end{figure}

STNMF can be seen as a type of method for clustering (see Methods). Similar to other clustering methods, the number of clusters, here modules $K$, is unknown at first. As $K$ is a free parameter, one has to choose $K$ before using STNMF. Similar to the previous study~\cite{Liu2017}, the number of meaningful subunit modules obtained by STNMF is not changed when $K$ is large enough, i.e., larger than the actual number of subunits used in the model (Fig.~\ref{fig:nmf}(A)). In other words, the result is convergent when using a large $K$, which is also seen by the convergence of Akaike information criterion when $K$ is larger~\cite{Onken_2016}. This advantage of STNMF, together with the constraint of the non-negativity condition, distinguishes STNMF from other traditional classification methods (see Methods).

With $K=4$, STNMF finds the exact number and structure of subunits. When $K=5$, the extra subunit in Fig.~\ref{fig:nmf}(A) is just noise with a low degree of coherence or auto-correlation in space. This signature can be used to determine the number of subunits when the actual number is unknown in real biological data~\cite{Liu2017}.

To test the robustness of STNMF, we applied some perturbations of the subunit structure to the model. The hypothesis is that the underlying computation is corresponding to the subunit structure. GC spiking responses are induced by the computation of subunits. If STNMF only reflects the properties of stimulus images, such as using NMF for face images~\cite{Lee_1999}, without taking into account the spiking computation, then the change of model subunits will not change STNMF subunit output. Instead, when STNMF can capture the underlying computation inside the network, one would expect that STNMF captures the change of the local structure of subunits. Therefore, we tested the hypothesis that subunits identified by STNMF are changed when the computation of the network is changed.

\begin{figure}[tht]
	\begin{center}
		\includegraphics[width=\columnwidth]{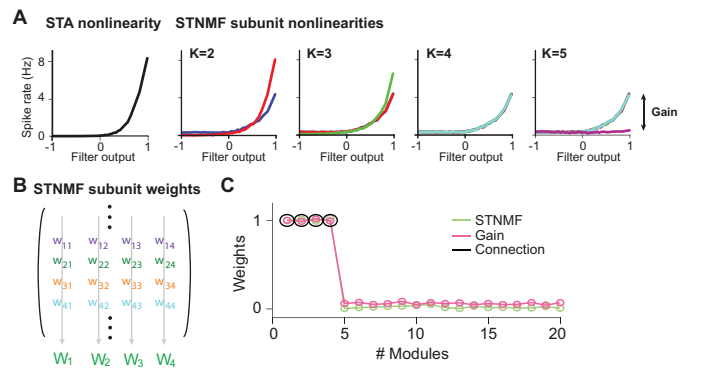}
	\end{center}
	\caption{ Subunit connection weight revealed by STNMF. (A)~Nonlinearities of each filter. (Left) Nonlinearity of STA filter. (Right) Nonlinearities of each STNMF subunits with $K=2,3,4,5$. Same color as in Fig.~\ref{fig:nmf}(A). (B)~STNMF subunit weights. Each column of the weight matrix is corresponding to a subunit. The sum of each column represents STNMF weight for each subunit. (C) STNMF weights (green), nonlinearity gains (purple) and connection weights (black) are identical. } \label{fig:nl}
\end{figure}

We manipulated the spatial structure of subunits as shown in Fig.~\ref{fig:nmf}(B, C). One perturbation is to have different sizes of subunits. Fig.~\ref{fig:nmf}(B) shows the case where the network consists of three subunits: two are in the same size, and one has a doubled size. By analyzing the spikes, a similar STA is obtained. However, STNMF precisely captures three subunits although they have different sizes. Another perturbation is to have an overlap between subunits. Fig.~\ref{fig:nmf}(C) shows that the network consists of three overlapping subunits. Similarly, the STA is a combination of all subunits. STNMF, on the other hand, can recover all three subunits separately.

In all the cases, the stimulus images are the same, but the spikes are different due to the changes of subunits and computations. Taken together, these results show that STNMF indeed captures the computation within the network, but not the pure effect of stimulus images.

After recovering the spatial subunit filter, one can also obtain the corresponding temporal filter for each subunit (Fig.~\ref{fig:tm1}). The temporal filter can be computed after the spatial filter is obtained by STNMF. A sequence of stimulus images is convolved by each spatial filter and then summed over all pixels to get a one-dimensional output. Spike-triggered analysis is then applied to this output to find the temporal filter.

Note that the temporal filters of subunits are not obtained by STNMF directly since here the effect of the time has been removed during the pre-processing of STNMF analysis. However, it is possible to obtain both spatial and temporal modules simultaneously (see \cite{Onken_2016}).
\begin{figure}[bt]
	\begin{center}
		\includegraphics[width=\columnwidth]{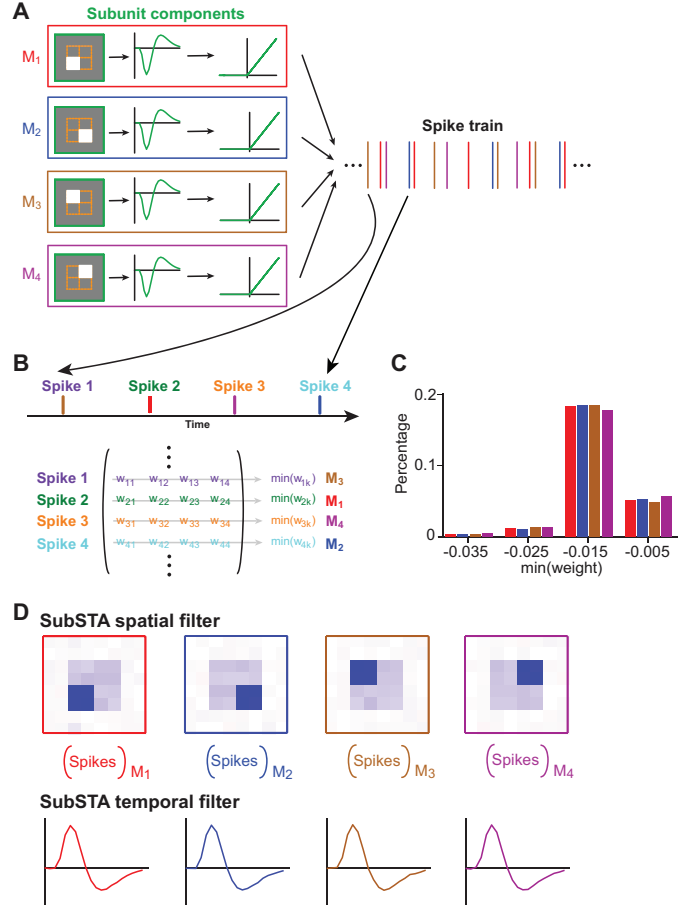}
	\end{center}
	\caption{  Classification of spikes by STNMF. (A)~Each spike can be seen as a contribution from one subunit $M_j$. (B)~Each spike is labeled with a corresponding subunit according to the minimal value of STNMF weight per row. Each column is corresponding to a subunit, and each row is corresponding to a spike. (C)~Histogram of weight minimums from four subsets of spikes showing a uniform distribution across four subunits. (D)~SubSTA spatial and temporal filter computed from one subset of spikes with STA analysis.
    } \label{fig:substa1}
\end{figure}

\subsection{Subunit connection weight revealed by STNMF}
Besides the spatial and temporal filter for each subunit, there is one last component in the model that needs to be identified: the connection weight of each subunit. For this purpose, we calculated the nonlinearity of each subunit by using its spatial and temporal filter and then averaged as a histogram mean~\cite{Chichilnisky2001a,Liu_2015}. To do so, stimuli are first convolved with the respective spatial and temporal filter to obtain a generator signal. This generator signal is then binned into 40 bins with variable bin size so that each bin contained the same number of data points. Then the nonlinearity is displayed as a histogram mean by plotting the average generator signal against the average spike rate for each bin.

Fig.~\ref{fig:nl}(A) shows that nonlinearities are changing with the parameter $K$. When $K=4$, all nonlinearities of four subunits are overlapping since the subunit connection weights used in the model are the same. Similarly, when $K=5$, a weak (flat) nonlinearity for the fifth subunit occurs due to the noise. The strength of nonlinearity can be characterized by the gain or the magnitude of the nonlinearity. The gain reflects how much the subunit contributes to spiking response, so it is closely related to subunit weight.

As STNMF is analyzing all the spikes, one expects that the gain can be revealed by STNMF. Indeed, we found this information can be extracted from the weight matrix of STNMF (Fig.~\ref{fig:nl}(B)). By averaging each column of the weight matrix, one can obtain a weight $W_j$ for each subunit $j$. Interestingly, the weight $W_j$ is identical to the connection weight of subunit $j$. All of these three measures, the STNMF weight, nonlinearity gain, and connection weight, are matched very well (Fig.~\ref{fig:nl}(C)). This indicates that the STNMF subunit weight provides a good estimate of the actual subunit connection weight. Such information is difficult to obtain in the biological data due to limitations in experimental techniques~\cite{Liu2017}.

\subsection{Classification of spikes by STNMF}
In the GC model, the final spikes are contributed by four subunits, thus, each spike could be induced by one subunit. Inspired by the clustering viewpoint of STNMF, one may wonder if STNMF can be used to classify all the GC spikes to four subsets of spikes such that each subset of spikes is mainly, if not completely, contributed by one specific subunit. 

\begin{figure}[bpt]
	\begin{center}
		\includegraphics[width=\columnwidth]{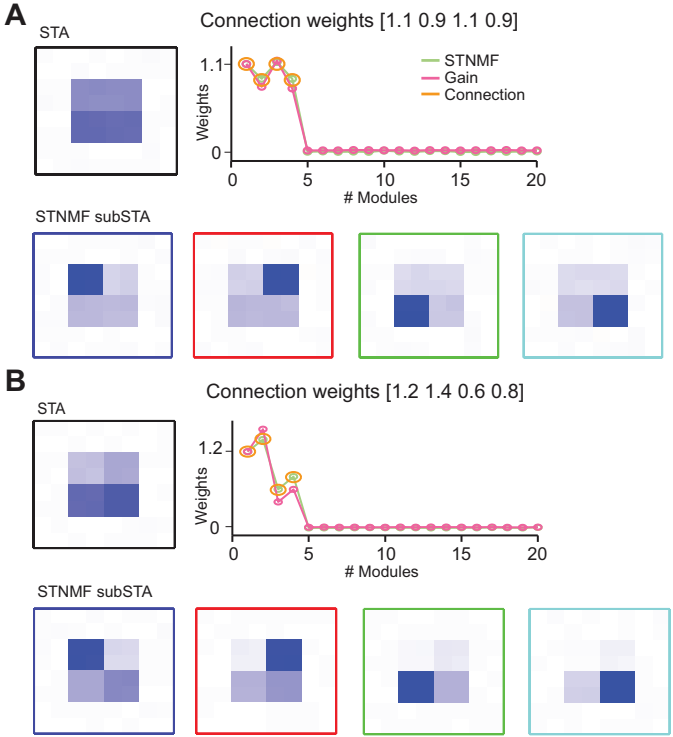}
	\end{center}
	\caption{ Effect of heterogeneous connection weights. (A)~Connection weights $[1.1, 0.9, 1.1, 0.9 ]$. (top, left) Spatial STA filter. (top, right) Module weights computed as normalized gain, NMF weight, and connection weight designed in the model. (bottom) SubSTA computed from subsets of spikes. (B)~Similar to (A) but with connection weights as $[1.2, 1.4, 0.6, 0.8]$.
    } \label{fig:weight}
\end{figure}

The STNMF weight matrix is subunit-specific for every column, but it is also spike-specific for every row. As each row corresponds to one individual spike, every spike can be labeled or classified according to one subunit as illustrated in Fig.~\ref{fig:substa1}(A). Note that the model is designed for OFF-type GC, and since the subunits are always non-negative, one can take the minimal value per row in weight matrix $w_{ij}$, for instance $\min(w_{1k}) = \min_j(w_{1j})$ for the first row, i.e., the first spike as in Fig.~\ref{fig:substa1}(B). The minimum index $j$ is this spike’s label for the subunit $k$. The value of the minimum weight can measure the contribution made by this subunit.

\begin{figure}[t!ht]
	\begin{center}
		\includegraphics[width=\columnwidth]{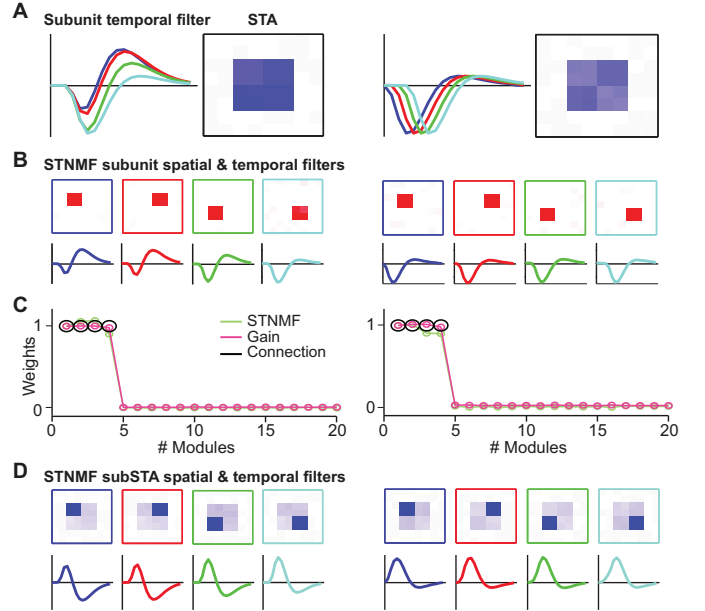}
	\end{center}
	\caption{ Effect of different temporal filters on STNMF results. (A)~Heterogeneous temporal filters with different amplitudes (left) and delays (right). Spatial STAs are similar. (B)~STNMF subunit spatial and temporal filter. Note both amplitude and delay changes are recovered by STNMF. (C)~Matching of STNMF weight, nonlinearity gain, and connection weight. (D)~Spatial and temporal filter of STNMF subSTA computed from each subset of spikes. Note here the polarity of STNMF subSTAs are opposite to STNMF subunits that are always positive. 
    } \label{fig:tm2}
\end{figure}

Now we can classify every spike into one specific $k$-th subunit $M_k$ since $k = min_j \{1,2,3,4\}$. For instance, for the first spike, Spike 1 is associated with the first row and the minimum value is at the 3rd column. Therefore, the first spike should be associated with the 3rd subunit $M_3$ with $k=3$. After doing this loop for all rows/spikes, we can label each spike with a specific subunit. For this particular model cell, we obtain four subsets of spikes for four subunits respectively. For every spike, there is a min(weight). The histogram of these min(weight) shows that these weights are indeed uniformly distributed across four subunits as in Fig.~\ref{fig:substa1}(C), meaning the connection weight of every subunit is the same as in the model.

Then for each subset of spikes, we computed the STA to get the spatial and temporal filter of each subunit as in Fig.~\ref{fig:substa1} (D). These spatial filters are similar to STNMF subunits. We name these filters as ``subSTAs''. 

As the subSTA reflects the contribution of one subunit to the GC spikes, one can test the robustness of subSTA by modifying the strength of the subunit connection. We manipulated the connection weights of four subunits as $[1.1, 0.9, 1.1, 0.9]$ and $[1.2, 1.4, 0.6, 0.8]$. Three measures, STNMF weights, nonlinearity gains and connection weights, are tightly matched as in Fig.~\ref{fig:weight}. As a result, the subSTAs computed from the subsets of spikes classified by STNMF are also faithfully similar to the subunits used in the model. 

During the pre-processing of data for STNMF, we obtained the ensemble of effective stimulus images where the temporal correlation was removed. This naturally poses the question whether all of the results obtained by STNMF could change when the temporal filters in the model change. In order to test this, we used two different perturbations for the temporal filter: different amplitudes and different delays between temporal filters (Fig.~\ref{fig:tm2}(A)). Similar to the previous study \cite{Liu2017}, the results are very robust. All of the properties of subunits, such as spatial and temporal filter (Fig.~\ref{fig:tm2}(B)), and matching of three measures, STNMF weight, nonlinearity gain, and connection weight (Fig.~\ref{fig:tm2}(C)), are robust. In addition, the subSTAs obtained by classified spikes are also faithfully reproduced. We also calculated temporal components that are the correspondence of spatial subSTAs. These temporal subSTAs are also closely matched to the subunit temporal filters designed in the model as in Fig.~\ref{fig:tm2}(D).

\subsection{Application of STNMF to real retinal data}

\begin{figure}[tb]
	\begin{center}
		\includegraphics[width=\columnwidth]{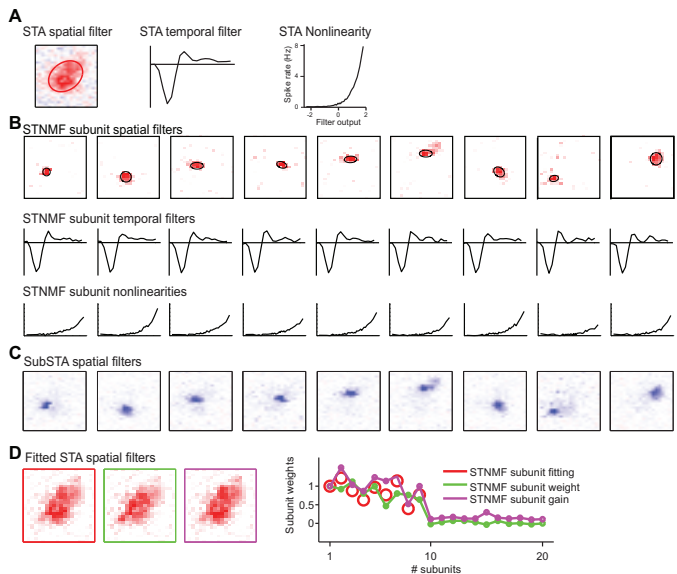}
	\end{center}
	\caption{ Overview of biological properties of a ganglion cell identified by STNMF. (A)~STA spatial filter, temporal filter, and nonlinearity. (B)~STNMF subunit spatial filters, temporal filters, and their nonlinearities. Circles are the outlines fitted with 2D Gaussian. (C)~SubSTA spatial filters from subsets of classified GC spikes. (D)~Reconstructed STA spatial filter by fitting with STNMF subunit (red), STNMF weights from weight matrix (green) and STNMF subunit gain from the nonlinearity (orange), respectively. These three measures are highly correlated (right). } \label{fig:substa2}
\end{figure}

We applied STNMF to the retinal GC data published previously~\cite{Liu2017,Onken_2016}. Briefly, the salamander retinal GCs were recorded with a multielectrode array under similar stimulation of spatiotemporal white noise as in the model above. An overview of application of STNMF to one GC is shown in Fig.~\ref{fig:substa2}, which is similar to what has been shown previously~\cite{Liu2017}. Standard spike-triggered analysis can get spatial receptive field, temporal filter, and nonlinearity as in Fig.~\ref{fig:substa2}(A). For this particular cell, STNMF can find nine subunits, i.e, BCs that connect to the current GC. The computational properties of these BCs include spatial receptive field, temporal filter, and nonlinearity as in Fig.~\ref{fig:substa2}(B). 

All the spikes from this GC can be further classified into nine subsets of spikes according to each subunit. Another STA analysis can get subSTA spatial filter for each subunit. Similar to the modeling result above, these subSTAs are highly similar to the subunit receptive fields identified by STNMF in Fig.~\ref{fig:substa2}(C).

\begin{figure}[tb]
	\begin{center}
		\includegraphics[width=\columnwidth]{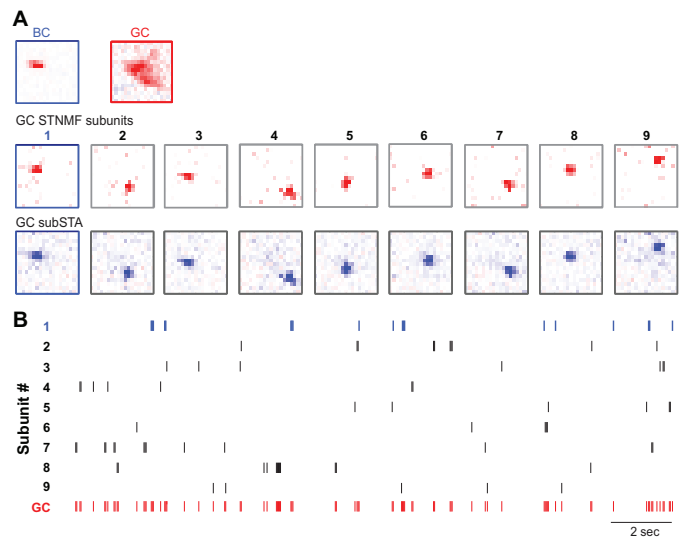}
	\end{center}
	\caption{ Subset of spikes contributed by BC. (A)~Receptive fields of BC and GC (top). STNMF subunit receptive fields (middle). SubSTA computed from subsets of classified GC spikes. 
(B)~Subsets of GC spikes classified by STNMF. 1st spike train is contributed by this BC together with the original GC spike train. BC data taken from Ref.~\cite{Liu2017}.
    } \label{fig:cc}
\end{figure}

Synaptic connection strength of each subunit can be computed in three different ways. 1) weights of fitting GC receptive field with all subunit receptive fields; 2) subunit weights calculated from the weight matrix of STNMF as above; 3) subunit gains calculated from each nonlinearity of subunit. Synaptic strengths from the last two measures can also be used to fit GC receptive field, which results in a similar picture of recovered GC receptive field (Fig.~\ref{fig:substa2}(D, left)). Although there is no ground-truth about the actual synaptic weights between BCs and this GC, all these three measures are highly correlated in Fig.~\ref{fig:substa2}(D, right). 

Once all GC spikes are classified into the subsets of BC spikes, one may test if these BC spikes are contributed by one actual bipolar cell. The dataset in \cite{Liu2017} does provide such a possibility as they simultaneously recorded one bipolar cell with a large population of ganglion cells. One example is shown in Fig.~\ref{fig:cc}(A) with the receptive fields of BC and GC. Again the subunits identified by STNMF and the subSTAs computed by subsets of subunit spikes are highly overlapping. The 1st subunit is highly overlapping with the recorded BC, which is shown in \cite{Liu2017} as an indication for actual connection between this BC and GC.

\begin{figure*}[tb]
	\begin{center}
		\includegraphics[width=1.9\columnwidth]{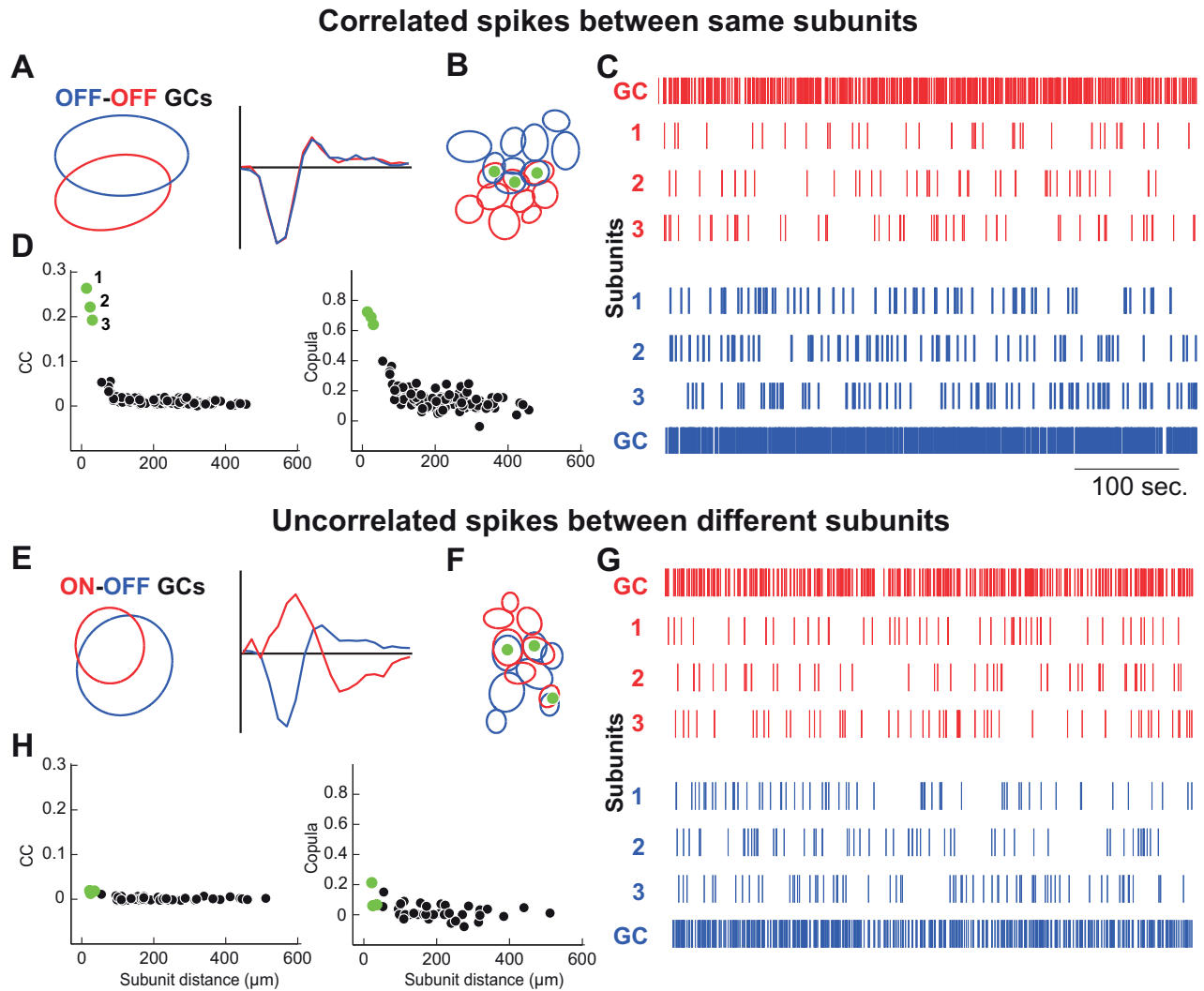}
	\end{center}
	\caption{ Subunit spikes induced by bipolar cells but not driven by shared stimuli. (A-D)~Correlated spikes between same subunits. (A)~Pair of OFF-OFF GCs with overlapping receptive fields and identical OFF-type temporal filters. (B)~Subunits obtained by STNMF for this pair of GCs. Three green pairs of subunits shared between these two GCs. (C)~Three subsets of spikes corresponding to three pairs of BCs colored in green in (B). Original GC spikes are shown at the top and bottom. (D)~Correlation coefficient (CC, left) and copula strength (right) for all pairs of BCs in (B). Top three numbered by 1, 2, and 3 are three subsets of spikes in (C). (E-H)~Similar to (A-D) but uncorrelated spikes between different subunits for a pair of BCs with one OFF and one ON type of GC. Three pairs of BCs colored in green are overlapping, but have different identity as ON and OFF BCs as indicated by their temporal filters in (E). Correlations between BCs are close to zero.    
    } \label{fig:overlap}
\end{figure*}

As now the spikes of GC can be classified into a subset of spikes for each BC in Fig.~\ref{fig:cc}(B), one can test if there is a functional connection between the 1st subunit and the recorded BC, besides that they are physically located at the same place. To explore this, we calculated correlation coefficient (CC) between the BC membrane potential and the subunit spike trains.
We found there is a stronger correlation between the 1st subunit and the BC (CC=0.14), compared to correlations of other subunits ( CC=$0.02 \pm 0.01, \textrm{mean} \pm \textrm{std}$).

The correlation coefficient is a measure of linear correlation which can miss non-linear correlations. To investigate whether there are any such non-linear couplings between the BC membrane potential and other subunits, we also looked at additional measures of dependence. The relationship between subunits and the BC membrane potential is complicated by the fact that the subunit activity consists of a discrete number of events (i.e. neuronal spikes), whereas the BC membrane potential is a continuous quantity. We analyzed this relationship calculating the spike-triggered average of the BC membrane potential and by binning the spike trains into short time windows (33~ms) and modeling the joint distribution of spike counts and membrane potential.

First, we calculated the spike-triggered average of the BC membrane potential and founds that it gives an amplitude for the 1st subunit as $1.35$ and other subunits as $0.14 \pm 0.08$.
Then, we also applied the vine copula with various parametric copula families
as a general statistical model of joint distributions that can represent couplings between discrete signals, here the subunit spike counts, and continuous signals, here the BC membrane potential. When the signals are mixed as discrete quantities or continuous quantities, vine copulas as models of the mixed joint distributions is quite useful~\cite{onken2016mixed}. These models include various choices for parametric bivariate copulas. Here, we use the Gaussian, student t and rotated Clayton copula families and use the canonical vine to extend these bivariate models to multivariate models~\cite{onken2016mixed}. We fit these models to the mixed data and use the copula parameters to quantify coupling strengths. For all copula families, a coupling strength of zero corresponds to independence. Here we found that the vine copula gives a coupling strength for the 1st subunit as $7.12$ and other subunits as $0.19 \pm 0.18$.
All these results indicate that the subset of spikes from the 1st subunit is indeed contributed by this BC.

A further investigation of coupling between subunits can be done at the population level. For each GC, there are a few subunits found by STNMF. One can look at pairs of two GCs, such that there are some overlapping subunits as illustrated by spatial receptive fields in Fig.~\ref{fig:overlap}(A), where one pair of GCs with the same type of fast OFF GCs \cite{Liu2017} is shown, together with another pair of GCs with the different cell types as fast OFF and ON cells. Such information about cell types can be seen by their temporal filters, where fast OFF GCs have identical filter shape, and ON GC has a positive polarity at its first peak in Fig.~\ref{fig:overlap}(B,F).

For each GC, a few subunits revealed by STNMF are shown in Fig.~\ref{fig:overlap}(C). There are considerable overlapping subunits in a pair of GGs. When subsets of spikes are obtained by STNMF, there is a possibility that correlation between spikes is induced by the overlapping spatial location, rather than produced by the same BC. Since spikes driven by the same stimulus inputs are highly correlated, in particular when spatiotemporal white noise stimuli are replaced by spatially uniform white noise~\cite{Liu_2015}, two trains of subunit spikes could be correlated when these two subunits are located in the same space. As a result, they are seeing the same stimuli at this spacial location. This possibility in our results can be examined by a population analysis of GCs. 

For the same type of GCs with overlapped subunits, the sample pair shown in Fig.~\ref{fig:overlap}(A) has three overlapping subunits in Fig.~\ref{fig:overlap}(B). Therefore, there are three shared subunit spike trains from each GC in this pair as in Fig.~\ref{fig:overlap}(C). We found that for each pair of overlapping subunits, their spike trains are highly correlated as characterized by correlation coefficient and vine copula coupling strength as in Fig.~\ref{fig:overlap}(D). However, the results obtained from different types of GCs are different as shown in Fig.~\ref{fig:overlap}(E-H). The sample pair shown in Fig.~\ref{fig:overlap}(E) has one OFF cell and one ON cell. This pair also shows highly overlapping subunits that have three spike trains classified by STNMF for each GC. In contrast to the pair of the same cell type, there is no correlation in subunit spike trains between overlapping subunits. A biological picture is that there are ON BCs in ON GC and OFF BCs in OFF GC. Although the overlapping ON and OFF BCs are located in a close-by spatial location due to the 3D structure of the retina, they are driven by the same stimulus but generate different spikes only when stimuli are presenting different parts: bright images for ON spikes v.s. dark images for OFF spikes. As a result, these two spike trains from ON and OFF BCs are not correlated. In other words, they are decorrelated due to the uncorrelated stimuli.

Taken together, these results show that correlations between subunit spikes are not driven by stimuli, but by the same BC identified by STNMF. This result confirms and extends the previous observation in \cite{Liu2017}, where the identity of BC was justified by subunit physical location only. Our results here go one step further to show that the identity of a BC can also be detected by means of functional properties, i.e., a subset of GC spikes contributed by this BC.




\section{Discussion}\label{discussion}
In this study, we proposed spike-triggered non-negative matrix factorization as a useful method for system identification of neuronal circuits. With a simple network model of the retinal ganglion cell with clearly defined subunit components, connections and weights, STNMF can be used to reveal all of these structural components within the network. Furthermore, STNMF can be used to classify the whole set of spikes of a ganglion cell into a few subsets of spikes, such that each subset of spikes is mainly contributed by one specific subunit. When applying STNMF to the retinal data, biological network components can be revealed. In particular, the classification of ganglion cell spikes shows that a subset of spikes is mainly contributed by one bipolar cell that connects to the target GC. 

Besides confirming what has been shown in the previous study~\cite{Liu2017}, where STNMF detected a layout of physical location of bipolar cells, here we significantly extended the power of the STNMF approach by analyzing the weight matrix given by STNMF. As a result, the STNMF weight matrix reflects functional properties of bipolar cells, including their synaptic connection weights and contributed spikes for the downstream ganglion cell. Therefore, STNMF is useful for uncovering the relevant functional and structural properties of neuronal circuits.

\subsection{Neuronal circuit at single cell level}
From the viewpoint of the postsynaptic neuron, properties of neuronal circuits revealed by STNMF include locations of presynaptic neurons and their synaptic connections and strengths/weights.

Structural components of a single postsynaptic neuron are revealed as a layout of presynaptic neurons. The organization of such a layout could be complex or simple. Depending on the type of neurons and animal species, the number of presynaptic neurons could be very large or small. For instance, in the cerebellum, Purkinje cells have a large dendritic tree with thousands of presynaptic connections, whereas unipolar cells have only one presynaptic fiber, and granular cells have an average of four presynaptic fibers~\cite{zampini2016mechanisms}. In salamander retina that was used in the current study, there are a few bipolar cells per ganglion cell~\cite{Liu2017}. 

Synaptic connections and weights are more difficult to identify. Traditionally, directly measuring these properties is established by pairwise (or triple and more) electrodes recording from pre- and post-synaptic neurons~\cite{Song2005,Jiang_2013}. Here we found that STNMF can directly identify these properties as part of the analysis of the simulated cell. A verification of this observation by experiment is possible for large-scale recordings of spiking and/or imaging of calcium signal activity of a population of neurons in the future, where inferring connections between neurons is feasible, for example, by means of graph theory or complex network analysis~\cite{Bassett_2017}.

\subsection{Classifying spikes of postsynaptic neurons}

A postsynaptic neuron receives a signal from a set of presynaptic neurons in multiple channels. Each presynaptic signal is ubiquitous in that the information from input to output is transformed in a nonlinear fashion. Such a nonlinearity is evidenced by the spiking activity of a neuron, where the incoming signal with mixed positive and negative signs is eventually transferred to a sequence of digital spikes. Such a feature becomes a fundamental principle of neuronal computation since the spiking mechanism was uncovered 60 years ago~\cite{hodgkin1952quantitative}.

STNMF implements the analysis of every single spike for one postsynaptic neuron. As a result, STNMF is naturally labeling every spike to one of the presynaptic neurons during the process of factorization. This relationship between spikes and presynaptic neurons is encoded in the weight matrix of STNMF. Here we decoded this information and classified all the spikes of a ganglion cell into a few subsets of spikes such that a subset of spikes is corresponding to one presynaptic neuron. In other words, these subsets of spikes are closely correlated to the activities of presynaptic neurons. 




Although the activity of a bipolar cell in the retina is traditionally viewed as a graded signal without spikes, it could still generate large deflections of the membrane potential that is similar to a spike event~\cite{Baden_2013}, we found that there exists a strong correlation between its membrane potential and the corresponding spikes. Both of these activities are generated by the stimulus of white noise checkers with a size of 30 $\mu m$. Therefore, it is possible, when stimuli are strong enough, to trigger strong activity in BC membrane potential that, in turn, can produce spiking activity in the connected GC. Indeed, one recent study found that one BC could trigger ganglion cell spiking under white noise bars stimulus by fitting a two-layer linear-nonlinear network similar to the model used in our current study~\cite{maheswaranathan2018inferring}. As for the other parts of the brain, the traditional view is that one presynaptic neuron may not be enough to drive a postsynaptic neuron to fire with a spike. However, a caveat here is that dendritic spikes could be larger than what we expected~\cite{Jason_2017}.

A simultaneous recording of both upstream BC and downstream GC in the retina is the ideal setting to test the identity of subunits revealed by STNMF~\cite{Liu2017}. Our current results go one step further to uncover the functional identity and potential contribution of BC for its downstream GC. Recent advancements in experimental techniques make it possible to record simultaneously the signals in the soma and multiple dendrites with both imaging and electrophysiology~\cite{ShimingTang_2017, RanDing_2017}. This protocol could provide an interesting test for the utility of STNMF.

\subsection{System identification of neural network  }

Retinal ganglion cells carry out visual computations from stimulus to response. One of the central problems in this computation is to find the encoding and decoding principles between stimulus and response~\cite{Yamins2016Using} for which a number of possible methods of system identification have been proposed in both visual neuroscience and computer vision~\cite{Yu2020, vance2018bioinspired, Marblestone_2016}. 

The input-output relation of sensory information has been traditionally modeled by some dynamic functions, for example, the Laguerre-Volterra model~\cite{Marmarelis_1993, Li_2011}, or trainable network models through unsupervised (e.g., spike-timing dependent plasticity)~\cite{Guyonneau_2004} or supervised learning~\cite{Yu_2016}. In contrast, detailed neuroscience knowledge provides a bottom-up approach with neural network models~\cite{Marblestone_2016, Hassabis2017}, whereas the underlying network structure needs to be cleverly designed by hand or selected from a massive pool of possible network architectures~\cite{Alexander_2018}. 

It has been observed in neuroscience experiments that specific features are encoded by specific neurons in the visual system, and also in other sensory systems~\cite{Gollisch_2010}. NMF itself can be viewed as a generative model~\cite{li2016advances}, whereas convolutional neural network is a supervised model. However, both types of methods can be used to extract the underlying features from data. Their potential usage for modeling input-output relations is evident: local structure features play an important role for computation~\cite{Liu2017, yan2017revealing, yan2018revealing}. Indeed, recent studies show that some NMF variants can go beyond shallow layered networks, like our modeled retina network with only two layers, to use a framework of deep architectures~\cite{Ahn2004, CICHOCKI_2007, LyuW13, Song_2013, Trigeorgis_2017} to learn a hierarchy of attributes of a given dataset. A combination of NMF and deep convolutional neural network holds promise to uncover hierarchical structures of neural networks~\cite{Tseng_2015, Geiger_2014, Williamson_2015, Kang_2015, Hui_Zhang_2016}. Therefore, further extensions of our current STNMF are likely to be fruitful for understanding the deep architecture of neuronal systems in the brain.

\section*{Acknowledgment}
We would like to thank Y. Zheng, Y. Zhang, L. He, Y. Yue and K. Du for helpful discussions.

\ifCLASSOPTIONcaptionsoff
  \newpage
\fi

\bibliographystyle{IEEEtran}
\bibliography{ref_all_liu}

\begin{thebibliography}{10}
\providecommand{\url}[1]{#1}
\csname url@samestyle\endcsname
\providecommand{\newblock}{\relax}
\providecommand{\bibinfo}[2]{#2}
\providecommand{\BIBentrySTDinterwordspacing}{\spaceskip=0pt\relax}
\providecommand{\BIBentryALTinterwordstretchfactor}{4}
\providecommand{\BIBentryALTinterwordspacing}{\spaceskip=\fontdimen2\font plus
\BIBentryALTinterwordstretchfactor\fontdimen3\font minus
  \fontdimen4\font\relax}
\providecommand{\BIBforeignlanguage}[2]{{%
\expandafter\ifx\csname l@#1\endcsname\relax
\typeout{** WARNING: IEEEtran.bst: No hyphenation pattern has been}%
\typeout{** loaded for the language `#1'. Using the pattern for}%
\typeout{** the default language instead.}%
\else
\language=\csname l@#1\endcsname
\fi
#2}}
\providecommand{\BIBdecl}{\relax}
\BIBdecl

\bibitem{helmstaedter2013connectomic}
M.~Helmstaedter, K.~L. Briggman, S.~C. Turaga, V.~Jain, H.~S. Seung, and
  W.~Denk, ``Connectomic reconstruction of the inner plexiform layer in the
  mouse retina,'' \emph{Nature}, vol. 500, no. 7461, pp. 168--174, 2013.

\bibitem{zeng2017neuronal}
H.~Zeng and J.~R. Sanes, ``Neuronal cell-type classification: challenges,
  opportunities and the path forward,'' \emph{Nature Reviews Neuroscience},
  vol.~18, no.~9, p. 530, 2017.

\bibitem{marc2013retinal}
R.~E. Marc, B.~W. Jones, C.~B. Watt, J.~R. Anderson, C.~Sigulinsky, and
  S.~Lauritzen, ``Retinal connectomics: towards complete, accurate networks,''
  \emph{Progress in Retinal and Eye Research}, vol.~37, pp. 141--162, 2013.

\bibitem{seung2014neuronal}
H.~S. Seung and U.~S{\"u}mb{\"u}l, ``Neuronal cell types and connectivity:
  lessons from the retina,'' \emph{Neuron}, vol.~83, no.~6, pp. 1262--1272,
  2014.

\bibitem{sanes2015types}
J.~R. Sanes and R.~H. Masland, ``The types of retinal ganglion cells: current
  status and implications for neuronal classification,'' \emph{Annual Review of
  Vision Science}, vol.~38, pp. 221--246, 2015.

\bibitem{demb2015functional}
J.~B. Demb and J.~H. Singer, ``Functional circuitry of the retina,''
  \emph{Annual Review of Vision Science}, vol.~1, pp. 263--289, 2015.

\bibitem{Chichilnisky2001a}
E.~J. Chichilnisky, ``A simple white noise analysis of neuronal light
  responses,'' \emph{Network}, vol.~12, no.~2, pp. 199--213, 2001.

\bibitem{Pillow2008Spatio}
J.~W. Pillow, J.~Shlens, L.~Paninski, A.~Sher, A.~M. Litke, E.~J. Chichilnisky,
  and E.~P. Simoncelli, ``Spatio-temporal correlations and visual signalling in
  a complete neuronal population,'' \emph{Nature}, vol. 454, no. 7207, p. 995,
  2008.

\bibitem{McFarland2013}
J.~M. McFarland, Y.~Cui, and D.~A. Butts, ``Inferring nonlinear neuronal
  computation based on physiologically plausible inputs,'' \emph{PLoS
  Computational Biology}, vol.~9, no.~7, p. e1003143, jul 2013.

\bibitem{Yu2020}
Z.~Yu, J.~K. Liu, S.~Jia, Y.~Zhang, Y.~Zheng, Y.~Tian, and T.~Huang, ``Toward
  the next generation of retinal neuroprosthesis: Visual computation with
  spikes,'' \emph{Engineering}, feb 2020.

\bibitem{Zhang2020}
Y.~Zhang, S.~Jia, Y.~Zheng, Z.~Yu, Y.~Tian, S.~Ma, T.~Huang, and J.~K. Liu,
  ``Reconstruction of natural visual scenes from neural spikes with deep neural
  networks,'' \emph{Neural Networks}, vol. 125, pp. 19--30, may 2020.

\bibitem{Song2005}
S.~Song, P.~J. Sj{\"o}str{\"o}m, M.~Reigl, S.~Nelson, and D.~B. Chklovskii,
  ``Highly nonrandom features of synaptic connectivity in local cortical
  circuits,'' \emph{{PLoS} Biology}, vol.~3, no.~3, pp. 507--519, 2005.

\bibitem{Jiang_2013}
X.~Jiang, S.~Shen, C.~R. Cadwell, P.~Berens, F.~Sinz, A.~S. Ecker, S.~Patel,
  and A.~S. Tolias, ``Principles of connectivity among morphologically defined
  cell types in adult neocortex,'' \emph{Science}, vol. 350, no. 6264, p.
  aac9462, 2015.

\bibitem{Liu2017}
J.~K. Liu, H.~M. Schreyer, A.~Onken, F.~Rozenblit, M.~H. Khani,
  V.~Krishnamoorthy, S.~Panzeri, and T.~Gollisch, ``Inference of neuronal
  functional circuitry with spike-triggered non-negative matrix
  factorization,'' \emph{Nature Communications}, vol.~8, no.~1, p. 149, jul
  2017.

\bibitem{Lee_1999}
D.~D. Lee and H.~S. Seung, ``Learning the parts of objects by non-negative
  matrix factorization,'' \emph{Nature}, vol. 401, no. 6755, pp. 788--791, oct
  1999.

\bibitem{Zhao_2016}
X.~Zhao, X.~Li, Z.~Zhang, C.~Shen, Y.~Zhuang, L.~Gao, and X.~Li, ``Scalable
  linear visual feature learning via online parallel nonnegative matrix
  factorization,'' \emph{{IEEE} Transactions on Neural Networks and Learning
  Systems}, vol.~27, no.~12, pp. 2628--2642, dec 2016.

\bibitem{ye2015multitask}
M.~Ye, Y.~Qian, and J.~Zhou, ``Multitask sparse nonnegative matrix
  factorization for joint spectral--spatial hyperspectral imagery denoising,''
  \emph{{IEEE} Transactions on Geoscience and Remote Sensing}, vol.~53, no.~5,
  pp. 2621--2639, 2015.

\bibitem{Kwon_2015}
K.~Kwon, J.~W. Shin, and N.~S. Kim, ``{NMF}-based speech enhancement using
  bases update,'' \emph{{IEEE} Signal Processing Letters}, vol.~22, no.~4, pp.
  450--454, apr 2015.

\bibitem{guan2012nenmf}
N.~Guan, D.~Tao, Z.~Luo, and B.~Yuan, ``{NeNMF}: an optimal gradient method for
  nonnegative matrix factorization,'' \emph{{IEEE} Transactions on Signal
  Processing}, vol.~60, no.~6, pp. 2882--2898, 2012.

\bibitem{gao2014machine}
B.~Gao, W.~L. Woo, and B.~W. Ling, ``Machine learning source separation using
  maximum a posteriori nonnegative matrix factorization,'' \emph{{IEEE}
  Transactions on Cybernetics}, vol.~44, no.~7, pp. 1169--1179, 2014.

\bibitem{pei2014automated}
X.~Pei, T.~Wu, and C.~Chen, ``Automated graph regularized projective
  nonnegative matrix factorization for document clustering,'' \emph{{IEEE}
  Transactions on Cybernetics}, vol.~44, no.~10, pp. 1821--1831, 2014.

\bibitem{blumensath2016directional}
T.~Blumensath, ``Directional clustering through matrix factorization,''
  \emph{{IEEE} Transactions on Neural Networks and Learning Systems}, vol.~27,
  no.~10, pp. 2095--2107, 2016.

\bibitem{wang2011fast}
H.~Wang, F.~Nie, H.~Huang, and F.~Makedon, ``Fast nonnegative matrix
  tri-factorization for large-scale data co-clustering,'' in \emph{IJCAI
  Proceedings-International Joint Conference on Artificial Intelligence},
  vol.~22, no.~1, 2011, p. 1553.

\bibitem{Devarajan_2008}
K.~Devarajan, ``Nonnegative matrix factorization: An analytical and
  interpretive tool in computational biology,'' \emph{PLoS Computational
  Biology}, vol.~4, no.~7, p. e1000029, jul 2008.

\bibitem{gold2011comparing}
K.~Gold, C.~Havasi, M.~Anderson, and K.~C. Arnold, ``Comparing matrix
  decomposition methods for meta-analysis and reconstruction of cognitive
  neuroscience results.'' in \emph{FLAIRS Conference}, 2011.

\bibitem{maruyama2014detecting}
R.~Maruyama, K.~Maeda, H.~Moroda, I.~Kato, M.~Inoue, H.~Miyakawa, and
  T.~Aonishi, ``Detecting cells using non-negative matrix factorization on
  calcium imaging data,'' \emph{Neural Networks}, vol.~55, pp. 11--19, 2014.

\bibitem{Beyeler_2016}
M.~Beyeler, N.~Dutt, and J.~L. Krichmar, ``3{D} visual response properties of
  mstd emerge from an efficient, sparse population code,'' \emph{Journal of
  Neuroscience}, vol.~36, no.~32, pp. 8399--8415, 2016.

\bibitem{pnevmatikakis2016simultaneous}
E.~A. Pnevmatikakis, D.~Soudry, Y.~Gao, T.~A. Machado, J.~Merel, D.~Pfau,
  T.~Reardon, Y.~Mu, C.~Lacefield, W.~Yang \emph{et~al.}, ``Simultaneous
  denoising, deconvolution, and demixing of calcium imaging data,''
  \emph{Neuron}, vol.~89, no.~2, pp. 285--299, 2016.

\bibitem{zhou2018efficient}
P.~Zhou, S.~L. Resendez, J.~Rodriguez-Romaguera, J.~C. Jimenez, S.~Q. Neufeld,
  A.~Giovannucci, J.~Friedrich, E.~A. Pnevmatikakis, G.~D. Stuber, R.~Hen
  \emph{et~al.}, ``Efficient and accurate extraction of in vivo calcium signals
  from microendoscopic video data,'' \emph{eLife}, vol.~7, p. e28728, 2018.

\bibitem{Hoyer_2004}
P.~O. Hoyer, ``Non-negative matrix factorization with sparseness constraints,''
  \emph{Journal of Machine Learning Research}, vol.~5, pp. 1457--1469, 2004.

\bibitem{Eggert_2004}
J.~Eggert and E.~Korner, ``Sparse coding and nmf,'' in \emph{IEEE International
  Joint Conference on Neural Networks}, vol.~4.\hskip 1em plus 0.5em minus
  0.4em\relax Institute of Electrical and Electronics Engineers ({IEEE}), 2004,
  pp. 2529--2533.

\bibitem{Olshausen1996}
B.~A. Olshausen and D.~J. Field, ``Emergence of simple-cell receptive field
  properties by learning a sparse code for natural images,'' \emph{Nature},
  vol. 381, no. 6583, pp. 607--9, 1996.

\bibitem{Hoyer_2003}
P.~O. Hoyer, ``Modeling receptive fields with non-negative sparse coding,''
  \emph{Neurocomputing}, vol. 52-54, pp. 547--552, jun 2003.

\bibitem{gollisch2008modeling}
T.~Gollisch and M.~Meister, ``Modeling convergent on and off pathways in the
  early visual system,'' \emph{Biological Cybernetics}, vol.~99, no. 4-5, pp.
  263--278, 2008.

\bibitem{vance2018bioinspired}
P.~J. Vance, G.~P. Das, D.~Kerr, S.~A. Coleman, T.~M. McGinnity, T.~Gollisch,
  and J.~K. Liu, ``Bioinspired approach to modeling retinal ganglion cells
  using system identification techniques,'' \emph{{IEEE} Transactions on Neural
  Networks and Learning Systems}, vol.~29, no.~5, pp. 1796--1808, 2018.

\bibitem{Sandler_2015}
R.~A. Sandler and V.~Z. Marmarelis, ``Understanding spike-triggered covariance
  using wiener theory for receptive field identification,'' \emph{Journal of
  Vision}, vol.~15, no.~9, p.~16, jul 2015.

\bibitem{Gauthier_2009}
J.~L. Gauthier, G.~D. Field, A.~Sher, M.~Greschner, J.~Shlens, A.~M. Litke, and
  E.~Chichilnisky, ``Receptive fields in primate retina are coordinated to
  sample visual space more uniformly,'' \emph{{PLoS} Biology}, vol.~7, no.~4,
  p. e1000063, apr 2009.

\bibitem{Ding_2010}
C.~H. Ding, T.~Li, and M.~I. Jordan, ``Convex and semi-nonnegative matrix
  factorizations,'' \emph{{IEEE} Transactions on Pattern Analysis and Machine
  Intelligence}, vol.~32, no.~1, pp. 45--55, jan 2010.

\bibitem{Kim_2007}
H.~Kim and H.~Park, ``Sparse non-negative matrix factorizations via alternating
  non-negativity-constrained least squares for microarray data analysis,''
  \emph{Bioinformatics}, vol.~23, no.~12, pp. 1495--1502, may 2007.

\bibitem{Li_2013}
Y.~Li and A.~Ngom, ``The non-negative matrix factorization toolbox for
  biological data mining,'' \emph{Source Code for Biology and Medicine},
  vol.~8, no.~1, p.~10, 2013.

\bibitem{Schwartz2006}
O.~Schwartz, J.~W. Pillow, N.~C. Rust, and E.~P. Simoncelli, ``Spike-triggered
  neural characterization,'' \emph{Journal of Vision}, vol.~6, no.~4, pp.
  484--507, 2006.

\bibitem{Onken_2016}
A.~Onken, J.~K. Liu, P.~C.~R. Karunasekara, I.~Delis, T.~Gollisch, and
  S.~Panzeri, ``Using matrix and tensor factorizations for the single-trial
  analysis of population spike trains,'' \emph{PLoS Computational Biology},
  vol.~12, no.~11, p. e1005189, nov 2016.

\bibitem{Liu_2015}
J.~K. Liu and T.~Gollisch, ``Spike-triggered covariance analysis reveals
  phenomenological diversity of contrast adaptation in the retina,'' \emph{PLoS
  Computational Biology}, vol.~11, no.~7, p. e1004425, jul 2015.

\bibitem{onken2016mixed}
A.~Onken and S.~Panzeri, ``Mixed vine copulas as joint models of spike counts
  and local field potentials,'' in \emph{Advances in Neural Information
  Processing Systems}, 2016, pp. 1325--1333.

\bibitem{zampini2016mechanisms}
V.~Zampini, J.~K. Liu, M.~A. Diana, P.~P. Maldonado, N.~Brunel, and
  S.~Dieudonn{\'e}, ``Mechanisms and functional roles of glutamatergic synapse
  diversity in a cerebellar circuit,'' \emph{eLife}, vol.~5, p. e15872, 2016.

\bibitem{Bassett_2017}
D.~S. Bassett and O.~Sporns, ``Network neuroscience,'' \emph{Nature
  neuroscience}, vol.~20, no.~3, p. 353, 2017.

\bibitem{hodgkin1952quantitative}
A.~L. Hodgkin and A.~F. Huxley, ``A quantitative description of membrane
  current and its application to conduction and excitation in nerve,''
  \emph{The Journal of Physiology}, vol. 117, no.~4, pp. 500--544, 1952.

\bibitem{Baden_2013}
T.~Baden, P.~Berens, M.~Bethge, and T.~Euler, ``Spikes in mammalian bipolar
  cells support temporal layering of the inner retina,'' \emph{Current
  Biology}, vol.~23, no.~1, pp. 48--52, 2013.

\bibitem{maheswaranathan2018inferring}
N.~Maheswaranathan, D.~B. Kastner, S.~A. Baccus, and S.~Ganguli, ``Inferring
  hidden structure in multilayered neural circuits,'' \emph{PLoS computational
  biology}, vol.~14, no.~8, p. e1006291, 2018.

\bibitem{Jason_2017}
J.~J. Moore, P.~M. Ravassard, D.~Ho, L.~Acharya, A.~L. Kees, C.~Vuong, and
  M.~R. Mehta, ``Dynamics of cortical dendritic membrane potential and spikes
  in freely behaving rats,'' \emph{Science}, vol. 355, no. 6331, p. eaaj1497,
  2017.

\bibitem{ShimingTang_2017}
M.~Li, F.~Liu, H.~Jiang, T.~S. Lee, and S.~Tang, ``Long-term two-photon imaging
  in awake macaque monkey,'' \emph{Neuron}, vol.~93, no.~5, pp. 1049--1057,
  2017.

\bibitem{RanDing_2017}
R.~Ding, X.~Liao, J.~Li, J.~Zhang, M.~Wang, Y.~Guang, H.~Qin, X.~Li, K.~Zhang,
  S.~Liang \emph{et~al.}, ``Targeted patching and dendritic ca 2+ imaging in
  nonhuman primate brain in vivo,'' \emph{Scientific reports}, vol.~7, no.~1,
  p. 2873, 2017.

\bibitem{Yamins2016Using}
D.~L.~K. Yamins and J.~J. Dicarlo, ``Using goal-driven deep learning models to
  understand sensory cortex,'' \emph{Nature Neuroscience}, vol.~19, no.~3, p.
  356, 2016.

\bibitem{Marblestone_2016}
A.~H. Marblestone, G.~Wayne, and K.~P. Kording, ``Toward an integration of deep
  learning and neuroscience,'' \emph{Frontiers in Computational Neuroscience},
  vol.~10, p.~94, sep 2016.

\bibitem{Marmarelis_1993}
V.~Z. Marmarelis, ``Identification of nonlinear biological systems using
  laguerre expansions of kernels,'' \emph{Annals of Biomedical Engineering},
  vol.~21, no.~6, pp. 573--589, nov 1993.

\bibitem{Li_2011}
W.~X. Li, R.~H. Chan, W.~Zhang, R.~C. Cheung, D.~Song, and T.~W. Berger,
  ``High-performance and scalable system architecture for the real-time
  estimation of generalized laguerre--volterra mimo model from neural
  population spiking activity,'' \emph{{IEEE} Journal on Emerging and Selected
  Topics in Circuits and Systems}, vol.~1, no.~4, pp. 489--501, dec 2011.

\bibitem{Guyonneau_2004}
R.~Guyonneau, R.~VanRullen, and S.~J. Thorpe, ``Temporal codes and sparse
  representations: A key to understanding rapid processing in the visual
  system,'' \emph{Journal of Physiology-Paris}, vol.~98, no. 4-6, pp. 487--497,
  jul 2004.

\bibitem{Yu_2016}
Q.~Yu, R.~Yan, H.~Tang, K.~C. Tan, and H.~Li, ``A spiking neural network system
  for robust sequence recognition,'' \emph{{IEEE} Transactions on Neural
  Networks and Learning Systems}, vol.~27, no.~3, pp. 621--635, mar 2016.

\bibitem{Hassabis2017}
H.~Demis, K.~Dharshan, S.~Christopher, and B.~Matthew, ``Neuroscience-inspired
  artificial intelligence,'' \emph{Neuron}, vol.~95, no.~2, pp. 245--258, jul
  2017.

\bibitem{Alexander_2018}
A.~J. Kell, D.~L. Yamins, E.~N. Shook, S.~V. Norman-Haignere, and J.~H.
  McDermott, ``A task-optimized neural network replicates human auditory
  behavior, predicts brain responses, and reveals a cortical processing
  hierarchy,'' \emph{Neuron}, vol.~98, no.~3, pp. 630--644, 2018.

\bibitem{Gollisch_2010}
T.~Gollisch and M.~Meister, ``Eye smarter than scientists believed: Neural
  computations in circuits of the retina,'' \emph{Neuron}, vol.~65, no.~2, pp.
  150--164, jan 2010.

\bibitem{li2016advances}
Y.~Li, ``Advances in multi-view matrix factorizations,'' in \emph{Neural
  Networks (IJCNN), 2016 International Joint Conference on Neural Networks},
  2016, pp. 3793--3800.

\bibitem{yan2017revealing}
Q.~Yan, Z.~Yu, F.~Chen, and J.~K. Liu, ``Revealing structure components of the
  retina by deep learning networks,'' \emph{NIPS Symposium on Interpretable
  Machine Learning. arXiv:1711.02837}, 2017.

\bibitem{yan2018revealing}
Q.~Yan, Y.~Zheng, S.~Jia, Y.~Zhang, Z.~Yu, F.~Chen, Y.~Tian, T.~Huang, and
  J.~K. Liu, ``Revealing fine structures of the retinal receptive field by deep
  learning networks,'' \emph{arXiv preprint arXiv:1811.0229}, 2018.

\bibitem{Ahn2004}
J.-H. Ahn, S.~Choi, and J.-H. Oh, ``A multiplicative up-propagation
  algorithm,'' in \emph{Proceedings of the twenty-first international
  conference on Machine learning}.\hskip 1em plus 0.5em minus 0.4em\relax
  Association for Computing Machinery, 2004, p.~3.

\bibitem{CICHOCKI_2007}
A.~Cichocki and R.~Zdunek, ``Multilayer nonnegative matrix factorization using
  projected gradient approaches,'' \emph{International Journal of Neural
  Systems}, vol.~17, no.~06, pp. 431--446, dec 2007.

\bibitem{LyuW13}
S.~Lyu and X.~Wang, ``On algorithms for sparse multi-factor nmf.'' in
  \emph{Advances in Neural Information Processing Systems}, 2013, pp. 602--610.

\bibitem{Song_2013}
H.~A. Song and S.-Y. Lee, ``Hierarchical data representation model-multi-layer
  {NMF},'' \emph{arXiv preprint arXiv:1301.6316}, 2013.

\bibitem{Trigeorgis_2017}
G.~Trigeorgis, K.~Bousmalis, S.~Zafeiriou, and B.~W. Schuller, ``A deep matrix
  factorization method for learning attribute representations,'' \emph{{IEEE}
  Transactions on Pattern Analysis and Machine Intelligence}, vol.~39, no.~3,
  pp. 417--429, mar 2017.

\bibitem{Tseng_2015}
H.-W. Tseng, M.~Hong, and Z.-Q. Luo, ``Combining sparse {NMF} with deep neural
  network: A new classification-based approach for speech enhancement,'' in
  \emph{2015 {IEEE} International Conference on Acoustics, Speech and Signal
  Processing ({ICASSP})}, 2015, pp. 2145--2149.

\bibitem{Geiger_2014}
J.~T. Geiger, F.~Weninger, J.~F. Gemmeke, M.~W{\"o}llmer, B.~Schuller, and
  G.~Rigoll, ``Memory-enhanced neural networks and {NMF} for robust {ASR},''
  \emph{IEEE/ACM Transactions on Audio, Speech and Language Processing
  (TASLP)}, vol.~22, no.~6, pp. 1037--1046, 2014.

\bibitem{Williamson_2015}
D.~S. Williamson, Y.~Wang, and D.~Wang, ``Deep neural networks for estimating
  speech model activations,'' in \emph{2015 {IEEE} International Conference on
  Acoustics, Speech and Signal Processing ({ICASSP})}.\hskip 1em plus 0.5em
  minus 0.4em\relax Institute of Electrical and Electronics Engineers ({IEEE}),
  apr 2015, pp. 5113--5117.

\bibitem{Kang_2015}
T.~G. Kang, K.~Kwon, J.~W. Shin, and N.~S. Kim, ``{NMF}-based target source
  separation using deep neural network,'' \emph{{IEEE} Signal Processing
  Letters}, vol.~22, no.~2, pp. 229--233, feb 2015.

\bibitem{Hui_Zhang_2016}
H.~Zhang, H.~Liu, R.~Song, and F.~Sun, ``Nonlinear non-negative matrix
  factorization using deep learning,'' in \emph{2016 International Joint
  Conference on Neural Networks ({IJCNN})}.\hskip 1em plus 0.5em minus
  0.4em\relax Institute of Electrical and Electronics Engineers ({IEEE}), jul
  2016, pp. 477--482.

\end{thebibliography}

\end{document}